# 2D materials coupled to hybrid metal-dielectric waveguides for THz technology

**P. Huang,[1] S. Massabeau,[1] J. Tignon,[2] S. Dhillon,[1] A. Degiron,[2] J. Mangeney[1,*]**

*[1] Laboratoire Pierre Aigrain, Département de physique de l'ENS, École normale supérieure, PSL Research University, Université Paris Diderot, Sorbonne Paris Cité, Sorbonne Universités, UPMC Univ. Paris 06, CNRS, 75005 Paris, France*
*[2] Centre de Nanosciences et de Nanotechnologies, CNRS, Univ. Paris-Sud, Université Paris-Saclay, France*
**Corresponding author: juliette.mangeney@lpa.ens.fr*

**In this letter, we propose hybrid metal-dielectric waveguides coupled to 2D materials that provide strong light-matter interaction at THz frequencies. We investigate the properties of the fundamental propagating modes and show that the strength of in-plane electric field components is maximized at the top of the dielectric strip on which the 2D material is deposited. Our simulation predicts 100 % modulation of THz light by tuning the Fermi level of a graphene sheet deposited onto a 1mm-long waveguide. We also show the potential of graphene multilayers coupled to these waveguides for achieving lasing at THz frequency. Our approach is compatible with CMOS or THz quantum cascade laser technologies.**

Graphene and other 2D layered materials such as black phosphorus and TMCD have attracted increasing attention for THz technology due to their unusual electrical and optical properties [1]. For instance, graphene can absorb THz photons, displays high electron mobility at room temperature, high optical nonlinearities [2] and tunable carrier densities. Likewise, black phosphorus is well suited for detection of THz radiation owing to its finite and direct bandgap, which provides high $I_{on}/I_{off}$ ratio, its huge carrier density tunability and large mobilities [3]. Also, efficient THz modulation has been demonstrated on $MoS_2$ on silicon substrate under optical excitation [4]. Moreover, the individual atomic planes of 2D materials can be mechanically separated from the bulk crystal and placed onto arbitrary substrates. As a result, 2D materials, which can realize many functions required for THz photonic circuits (e.g., modulation and detection of photons), can be easily integrated with other components based on silicon technologies or THz quantum cascade laser technologies. However, the inherent thinness of 2D materials severely limits their interaction with normal incident light.

In recent years, several strategies have been proposed to enhance the interaction between THz optical field and the 2D material. One approach is to integrate 2D materials with photonic structures or into optical cavities. G. Liang *et al.* demonstrate 100 % absorption of the THz light by a graphene sheet incorporated into the cavity of a quantum cascade laser albeit with a bandwidth limited to the linewidth of the cavity resonance [5]. Alternatively, 2D materials have been integrated with optical waveguides to enhance the light–matter interaction in a non-resonant manner. Locatelli *et al.* [6] and M. Mittendorff *et al.* [7] propose advanced dielectric waveguiding structures based on a coupler with a graphene sheet between two waveguides and a graphene layer in a middle of a silicon waveguide, respectively. The latter enables to reach 90% of absorption from 0.2 to 0.7 THz for a 10 mm-long waveguide. Besides, V. Ryzhii *et al.* have investigated graphene multilayers integrated in metal slot-line waveguides, dielectric waveguides [8] and surface plasmonic metal waveguides [9]; they demonstrated the high potential of these components for THz lasing at room temperature.

Here, we propose hybrid metal-dielectric waveguides coupled to 2D materials deposited on top that are easy to fabricate and provide strong light-matter interaction at THz frequencies. The structure of the hybrid metal-dielectric waveguides is depicted in Fig.1. It consists of a continuous 500 nm thick metal film and a dielectric strip made of high resistivity Si (typ. n=7.5 $10^{11}$cm$^{-3}$) or low-doped GaAs (typ. n=$10^{14}$ cm$^{-3}$) layers. The propagating electric field is guided within the dielectric strip of width $w$ and height $h$. A 300 nm thick $SiO_2$ layer and the 2D material are deposited on the top of the hybrid metal-dielectric waveguide. The $SiO_2$ layer allows for electrostatic modulation of the 2D material. Two electrical contacts are deposited on the 2D material and the metal layer is used to apply a gate voltage. Applying a gate voltage between the 2D material and the metal film changes the carrier concentration in the 2D material sheet and therefore its chemical potential. 2D materials coupled to these hybrid metal-dielectric waveguides could provide key components for advanced on-chip THz systems such as sources, modulators or detectors. The use of silicon as the dielectric strip material has the advantage to be compatible with CMOS technology whereas the use of GaAs is attractive for monolithically integration with metal-metal THz quantum cascade lasers [10].

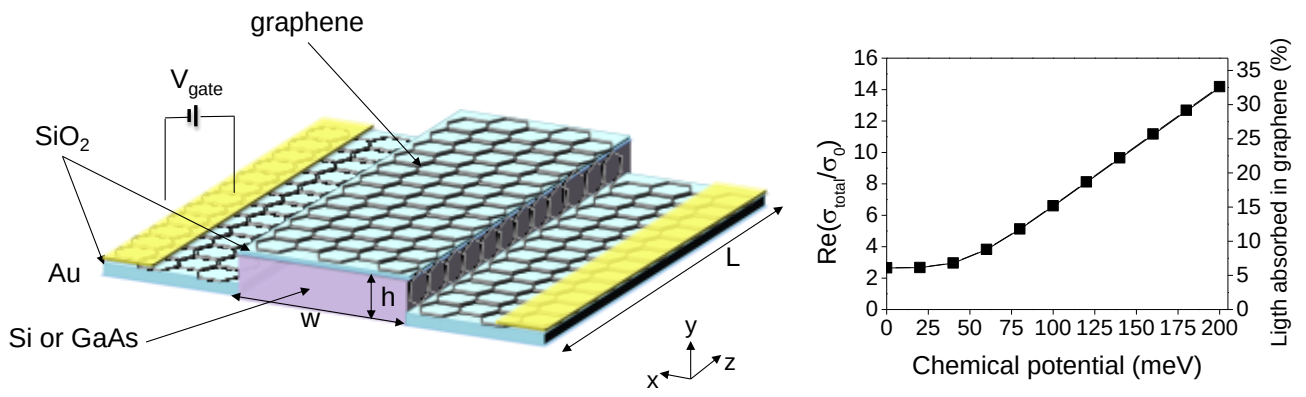

Fig. 1 Right: Sketch of the 2D material coupled to the waveguide structure. Left: Calculated real part of the total conductivity and the light absorbed in a single layer graphene for normal incident light illumination (without waveguide geometry) as a function of μ at 1 THz.

For the numerical analysis of these hybrid metal-dielectric waveguides, we assume that they are infinitely long along the propagation direction z so that the solutions can be written as $E = E_0(x,y)e^{in_{eff}k_0 z}$, where $E_0$ is the electric field in the transverse xy plane, $n_{eff}$ is the complex effective index and $n_{eff}k_0 = n_{eff}2\pi$⍰⍰ is the wavevector of the mode. Using commercial finite element code, we then solve for the eigen-modes of the system. The conducting properties of Au are modeled by using a complex permittivity fitting the experimental values compiled in Ref [11]. The 2D material is incorporated as a boundary layer between two regions of dielectrics with surface conductance, described by $\vec{n} \times (\overrightarrow{H_1} - \overrightarrow{H_2}) = \overrightarrow{J_s} = \sigma\vec{E}$ [12]. Here, $\vec{n}$ is the surface normal unit vector, σ is the complex conductivity of the 2D material, $\overrightarrow{H_1}$ and $\overrightarrow{H_2}$ are the magnetic fields on the opposite sides of the boundary and $\vec{E}$ the electric field at the boundary.

We first investigate the fundamental quasi-TE and quasi-TM modes propagating along these hybrid metal-dielectric waveguides (without 2D material on the top). We consider a silicon strip with a width *w*=300 μm and a height h=30 μm. Figure 2a show the spatial distribution of the two dominant electric field components, $E_x$ and $E_z$, for the quasi-TE mode ($|E_z| \ll |E_x\vec{x} + E_y\vec{y}|$). Both components can interact with a 2D layer deposited on the strip. The strength of the electric field $E_x$ is the highest and the longitudinal electric field $E_z$ reaches at maximum $0.374E_x$. The third component $E_y$ remains 15 dB weaker. Owing to the metal layer that acts as a reflector, the strength of $E_x$ and $E_z$ are maximized at the antinode of the optical field, localized in our geometry at the air/dielectric interface. For the quasi-TM mode ($|H_z| \ll |H_x\vec{x} + H_y\vec{y}|$), the two dominant electric field components shown on Fig.2b are $E_y$ and $E_z$ and thus, only the longitudinal component $E_z$ can interact with a 2D material deposited on the top of the strip. The electric field strength is nearly equally distributed between $E_y$ and $E_z$ since the maximum of $E_z$ reaches 90 % of the maximum of $E_y$. The strength of the in-plane electric field component $E_z$ is highest at the dielectric/air interface. Thus, these quasi-TE and quasi-TM modes are very promising for strong light-matter interaction with a 2D material deposited on this interface.

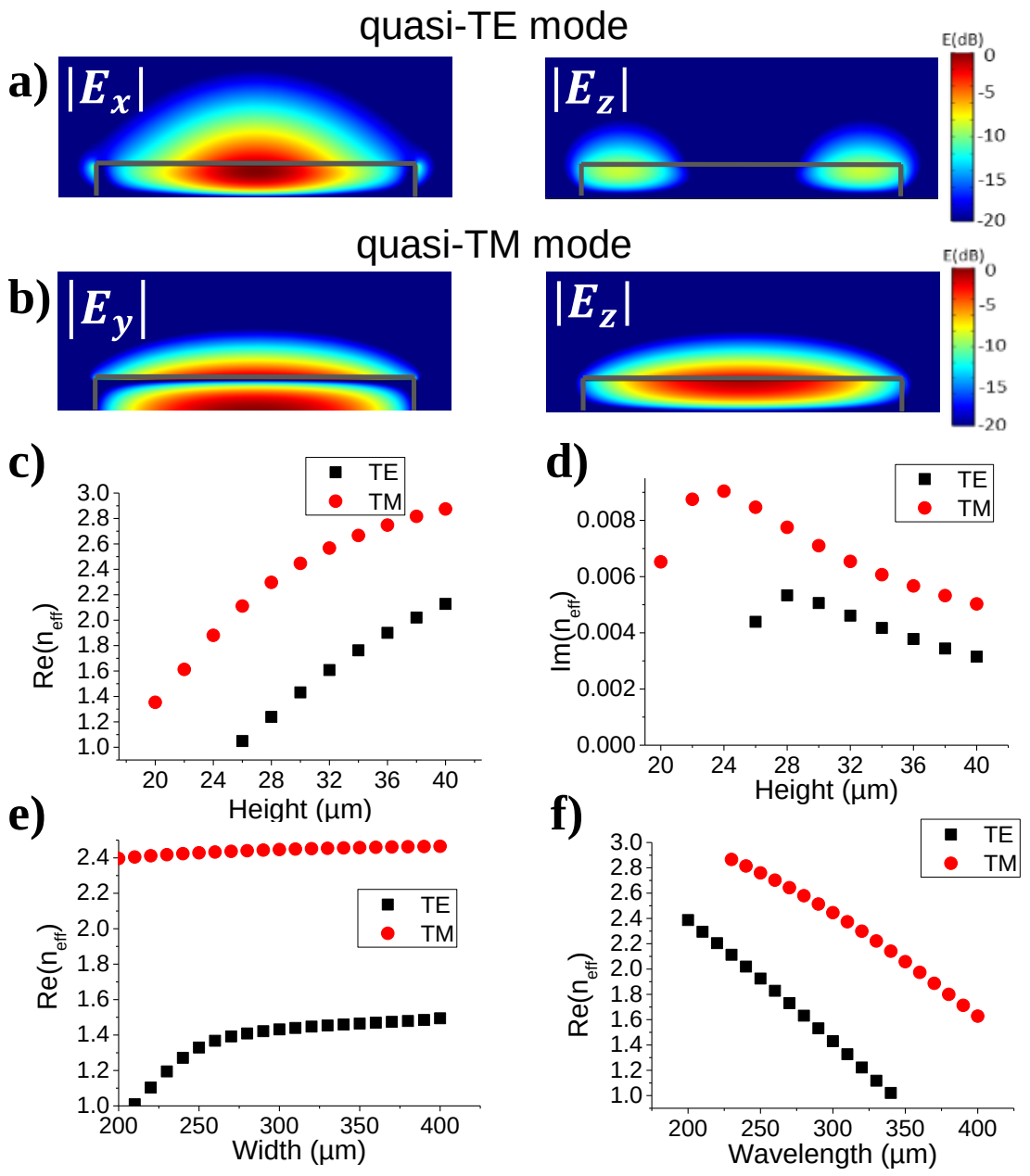

Fig. 2 Calculated electric field components in x and z directions of the fundamental quasi-TE mode (a) and in y and z directions of the fundamental quasi-TM mode (b) for an hybrid metal-silicon waveguide with a strip height of 30 μm and a strip width of 300 μm at a frequency of 1 THz. Real (c) and imaginary parts (d) of the effective index $n_{eff_WG}$ of quasi-TE and quasi TM modes as a function of the stripe height h (w=300μm) at 1 THz. e) Real part of the effective index $n_{eff_WG}$ of quasi-TE and quasi TM modes as a function of the strip width w (h=30μm) at 1 THz. f) Dispersion relation of the modes from 200 μm (1.5 THz) to 400 μm (0.75 THz) for metal-silicon waveguide with a strip height of 30 μm and a width of 300 μm.

To go further, we examine how the mode properties depend on the dimensions of the strip waveguide. Figure c-f report the evolution of the effective mode index, $n_{eff_WG}$, as a function of the width, the height and the wavelength respectively for the two fundamental modes. A cut-off of the quasi-TE mode is clearly observed in the evolution of the real and imaginary part of the mode effective index as a function of the width, the height and wavelength. Indeed, for small dimensions of the strip, such as h<λ/(4$n_{eff_WG}$) (Fig. 2c,d) or w<λ/$n_{eff_WG}$ (Fig. 2e), the quasi-TE mode is no longer a guided mode. In return, for large dimensions of the strip, the quasi-TE mode is well guided and $n_{eff_WG}$ saturates. A cut-off of the quasi-TM mode is also observed as the dimensions of the stripe are reduced. For h<λ/(4$n_{eff}$), the mode is no longer a guided mode. The imaginary part of the mode effective index for both quasi-TE and quasi-TM modes is due to free-carrier absorption of the electric field components that penetrate in the metal layer. This effect results in propagation losses that are described by a linear absorption coefficient given by $\alpha_{WG} = -4\pi Im(n_{eff_{WG}})/\lambda$. These propagation losses $\alpha_{WG}$ remain smaller than 1.1 dB/mm and 1.9 dB/mm for quasi-TE and quasi-TM modes respectively over a large range of stripe dimensions. The dispersion relation of the quasi-TE and quasi-TM modes, shown in Fig.2f from 0.75 (400 µm) to 1.5 THz (200 µm), highlight that these hybrid metal-dielectric waveguides are relatively broadband, making them interesting for pulsed THz excitation. This analysis shows that the quasi-TE mode is reminiscent of the quasi-TE mode of a rectangular dielectric waveguide surrounded by air [13], except that it is perturbed by the metallic layer that pushes up to the dielectric/air interface the electric field components tangent to the metallic layer. On the other hand, the quasi-TM mode has all the characteristics of a Goubau mode; it is bound to a finite conductivity metal stripe coated with a thin dielectric layer, the longitudinal field component is maximal at the dielectric/air interface and the mode shows a cut-off for height h<λ/(4$n_{eff}$) [14].

We now focus on the light-matter coupling between THz photons and single graphene layers integrated onto the hybrid metal-dielectric waveguides. We assume that the chemical potential of the graphene layer, µ, can be tuned by applying an appropriate gate voltage from 0 to 200 meV, as shown in Fig. 1. The total complex conductivity of the graphene sheet, governed by both intraband and interband processes in this spectral region, is calculated using $\sigma(\omega) = \sigma_{inter}(\omega) + \sigma_{intra}(\omega)$ [15]. The intraband conductivity is given by $\sigma_{intra}(\omega) = D\tau/\pi\,(1 - i\omega\tau)$ where $D = \frac{2e^2}{\hbar^2} k_B T ln[2\cosh(\mu/k_B T)]$ is the energy-dependent density of states in graphene, $\tau$ is the electron momentum scattering time. For our calculations, we assumed a carrier relaxation time of τ = 40 fs. The interband conductivity is described by: $\sigma_{inter} = \frac{\pi e^2}{4h}\left[\tanh\left(\frac{\hbar\omega+2\mu}{4k_B T}\right) + \tanh\left(\frac{\hbar\omega-2\mu}{4k_B T}\right)\right]$. In Fig. 1b is plotted the real part of the total conductivity and the absorption rate for normal incident light illumination of a single graphene sheet (without waveguide geometry) as a function of µ.

We first consider a hybrid metal-silicon waveguide with a height of 30 µm and a width of 300 µm with effective index mode $n_{eff_WG}$=1.432+i0.005 and $n_{eff_WG}$=2.4452+i0.007 for quasi-TE mode and quasi-TM mode respectively. In the presence of a single graphene layer on the top of the dielectric strip, the real part and imaginary parts of the effective mode index $n_{eff}$ increases. Figure 3a,b show the slight increase of Re($n_{eff}$) as the chemical potential is increased whereas Im($n_{eff}$) increases significantly. As expected, the quasi-TE mode is more affected by the graphene layer than the quasi-TM mode. Figure 3b shows the absorption of the graphene layer coupled to the waveguide calculated using $\alpha = 4\pi\, Im(n_{eff})/\lambda$. It reaches up to 24.7 dB/mm and 15.6 dB/mm at µ=200 meV for quasi-TE and quasi-TM mode respectively. Since this absorption is proportional to $Re(\sigma_{graphene})\Gamma$ with Γ τhe overlap factor between the single graphene layer and the mode field, these results demonstrate the enhancement of the light-matter interaction provided by our waveguide geometry. The quasi-TM mode has a significant longitudinal component, which could not be achieved with usual dielectric waveguides, providing a significant absorption of this quasi-TM mode by the graphene layer making these waveguides also very attractive for modulating the TM polarized light. As an illustration, we calculate the attenuation of hybrid metal-GaAs waveguides at 2.5 THz and 4 THz frequencies, and the corresponding depth modulation. The modulation depth, which is calculated between two different chemical potential states using $(T_{\mu=0meV} - T_{\mu=200meV})/T_{\mu=0meV}$ where T is the transmission coefficient, is a key parameter in view of light modulation applications. With the current waveguide designs, attenuation of 49 dB/mm and 34.7dB/mm can be obtained at 2.5 THz for quasi-TE and quasi-TM modes respectively as shown on Fig. 2c (plain and dashed black lines are the attenuation of the waveguide without graphene for quasi-TE and quasi-TM modes respectively). Figure 2d demonstrates that full modulation can be achieved in a single graphene layer coupled to hybrid metal-dielectric waveguides of 1mm length for both modes and for several waveguide geometries at 1 THz (metal-Si waveguide, h=30 µm, w=300µm), 2.5 THz (metal-GaAs waveguide, h=11 µm, w=120 µm), and 4 THz (metal-GaAs waveguide, h=7.5 µm, w=94 mm). Since the quasi-TM mode well matches the emission mode of metal-metal THz quantum cascade lasers [16], single graphene layer coupled to metal-GaAs waveguide is very attractive for light modulation of THz quantum cascade lasers. Moreover, our metal-dielectric waveguide platform has the advantage to be fully compatible with well-mastered wafer bonding processes, as it doesn't require unconventional fabrication steps to bury the 2D layer within the guiding structure. This study focuses on graphene layer but could be extend to other 2D materials such as doped black phosphorus and doped TMD materials such as $MoS_2$ and $WSe_2$ that also possess intraband conductivity in the THz range.

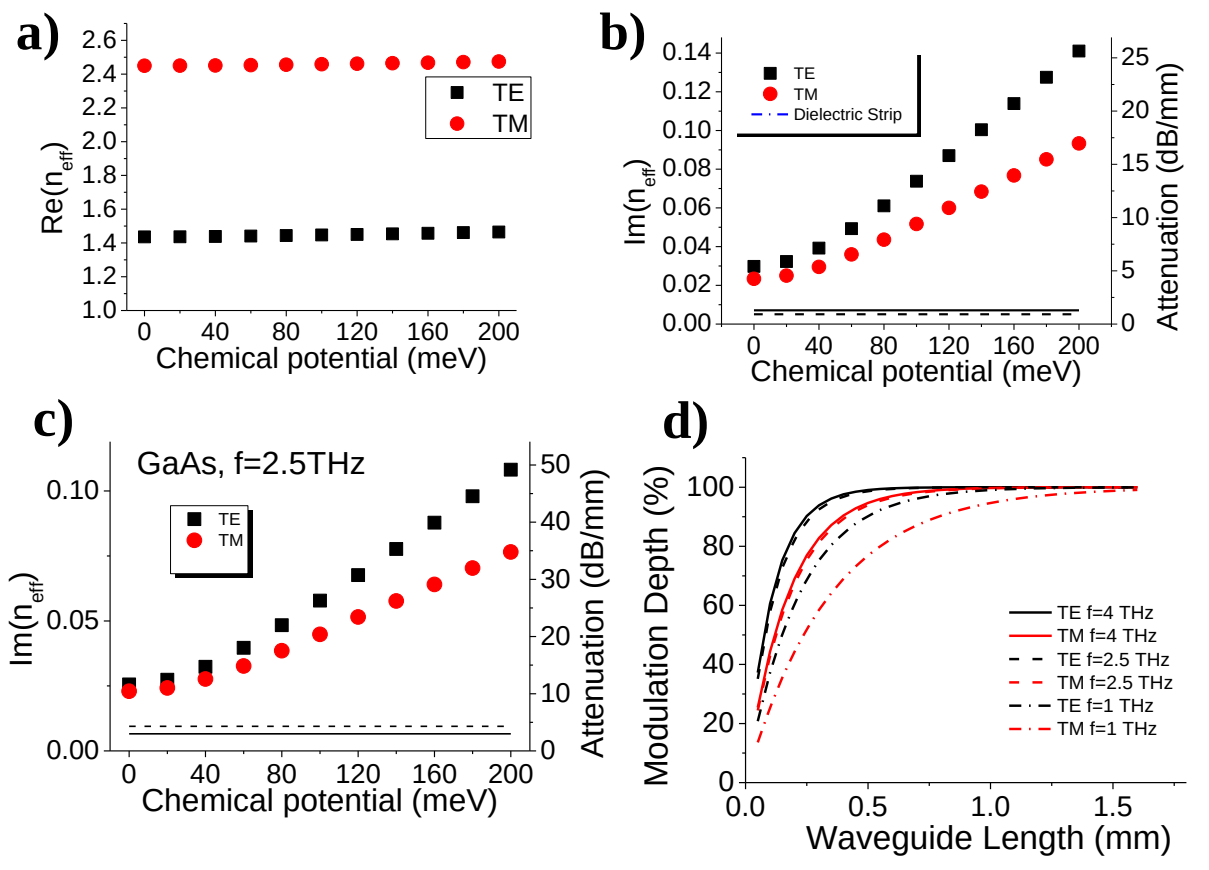

Fig. 3 Real (a) and imaginary (b) parts of the effective index mode as a function of the chemical potential of the single graphene layer coupled to a metal-Si waveguide (h=30µm, w=300µm) at 1 THz for quasi-TE and quasi- TM modes. c) Imaginary parts of the effective index mode as a function of the chemical potential of the single graphene layer coupled to a metal-GaAs waveguide (h=11µm, w=120µm) at 2.5 THz for both modes. Plain and dashed black lines are the attenuation of the waveguide without graphene for quasi-TE and quasi-TM modes respectively. d) Calculated modulation depth for different waveguide geometries as a function of the waveguide length L for quasi-TE and quasi-TM modes.

For very low chemical potential, the conductivity of 2D materials is mainly governed by interband processes at THz frequencies, which have no equivalent in usual semiconductor materials. Indeed, interband transitions are allowed at THz frequencies in graphene owing to its gapless [17] and also in n-doped black phosphorus by applying a perpendicular electric field [18] or gate electrode [19]. Interband processes at THz frequencies can lead to THz optical gain in population-inverted graphene and are also exploited for carrier generation in THz detectors based on photovoltaic effect. For these applications, it is crucial to enhance the interband absorption, limited to only 2.3% when light passes through a monolayer of graphene, see Fig. 1b. We now focus on this interband regime obtained for $|\mu| < \hbar\omega/2$ (i.e. µ<5 meV for photons at 2.5 THz). We assume low temperature range (~10K) so that the interband conductivity equals to the universal value $\sigma_0 = e^2/4\hbar = 6.08x10^{-5}$S. Figure 4 (right) shows the light absorbed by interband process at 2.5 THz and 4 THz as a function of the length of the waveguide for both quasi-TE and quasi-TM modes. We consider stripes made of high-resistivity silicon at 2.5 THz (compatible with silicon technology) and of low-doped GaAs at 4 THz (compatible with THz quantum cascade laser technology). We observed that more than 90% of the THz light is absorbed for waveguide length of 3 mm for all configurations, demonstrating clearly the enhancement of light-matter interaction provided by the waveguide geometry. The interband modal absorption reaches 18.5 cm$^{-1}$ and 16.5 cm$^{-1}$ for quasi-TE and quasi-TM modes respectively at 2.5 THz. These values show the high potential of graphene coupled metal-Si waveguide for photodetectors based on photovoltaic effect. The frequency of 4 THz is attractive for THz amplification since it is higher than the minimum frequency at which the optical gain can be achieved in population-inverted graphene at low temperature, due absorption assisted by the carrier-carrier scattering [20]. Figure 4 (left) presents the interband modal absorption at this frequency as a function of the number of independent graphene sheets deposited on the top of the hybrid metal-GaAs waveguide (h=7.5 µm, w=94 mm). Due to intraband absorption that effectively competes with interband amplification, the maximum achievable optical gain in population-inverted graphene is limited to typically $-0.75\alpha_{interband}$. The latter value has to dominate the total losses for achieving THz lasing. The dashed lines in Fig. 4 represents the total losses $\alpha_{total} = \alpha_{WG} + \alpha_m$ for both modes for a waveguide length of 2mm. The waveguide absorption loss ⍰$_{WG}$ equal to 15.3cm$^{-1}$ and 21.9 cm$^{-1}$ for quasi-TE and quasi-TM mode respectively and the mirror loss ⍰$_m$ due to the finite mirror reflectivity, defined by $\alpha_m = -\ln(R)/L$ with L the waveguide length and $R = ((n_{eff} - 1)/(n_{eff} + 1))^2$ , equals to 17.6 cm$^{-1}$ and 9 cm$^{-1}$ for quasi-TE and quasi-TM mode respectively. We observe that inserting 3-4 layers of graphene within the waveguide structure can provide enough optical gain for interband THz lasing. Note that at room temperature, the interband conductivity is lowered due to thermal carrier distribution and the number of graphene sheets required for achieving interband THz lasing is higher.

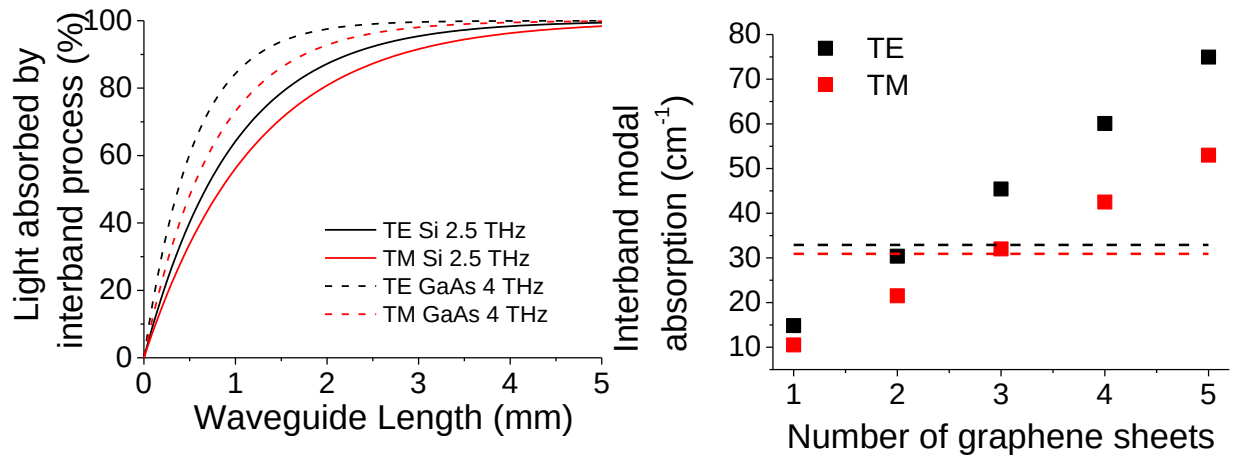

Fig. 4 Right: Interband light absorption in a single graphene sheet coupled to hybrid metal-dielectric waveguides as a function of L. Left: Calculated interband modal absorption at 4 THz as a function of the number of independent graphene sheets of an Au-GaAs hybrid waveguide (h=7.5 µm, w=94 µm). The dashed black and red lines represent the total losses for TE and TM modes respectively for a waveguide length of 2 mm.

In conclusion, we present an original hybrid metal-dielectric waveguide that provides high interaction between 2D materials deposited on top and THz photons. We investigate the fundamental quasi-TE and quasi-TM modes propagating along these hybrid metal-dielectric waveguides and show that, owing to the high reflectivity and low loss of metals at THz frequencies, the strength of in-plane electric field components of the propagating modes is maximized at the top of the dielectric strip on which the 2D material is deposited. Our simulation predict 100 % modulation of the THz light by tuning the Fermi level of a graphene sheet coupled to 1mm-long hybrid metal-dielectric waveguides, that is very attractive for THz modulation applications. We also show the potential of graphene multilayers under interband regime coupled to these hybrid metal dielectric waveguides for achieving lasing at THz frequencies. Our approach relies for fabrication on well-mastered wafer bonding techniques, does not require incorporating the 2D materials into the core of a dielectric waveguide, and is compatible with CMOS technology or THz quantum cascade lasers.

**Funding.** French National Agency (ANR-17)

## References.